\documentclass[conference]{IEEEtran}

\usepackage{tabu}
\usepackage[nolist]{acronym}
\usepackage{multirow}
\usepackage{graphicx}
\usepackage{amsmath}
\usepackage{mathrsfs}
\usepackage{amsfonts}
\usepackage[caption=false,font=footnotesize]{subfig}
\usepackage{color}
\usepackage{siunitx}
\usepackage{xcolor}

\usepackage{enumitem}

\usepackage{hyperref}
\usepackage{upgreek}
\usepackage{commath} %for abs value and norm

\definecolor{orange}{RGB}{255, 165, 0}

\def\CN{\mathcal{N}_{\mathbb{C}}} %Complex Gaussian distribution
\newcommand{\vect}[1]{\mathbf{#1}}

\def\taupu{\tau_{p}} %Uplink pilots
\def\bphiu{\boldsymbol{\phi}} %Notation of uplink pilot sequences
\def\Htran{\mbox{\tiny $\mathrm{H}$}}

\acrodef{ARP}{Access Reservation Protocol}
\acrodef{UE}{User Entity}
\acrodef{RAO}{Random Access Opportunity}
\acrodef{MTC}{Machine Type Communication}
\acrodef{TTI}{Transmission Time Interval}
\acrodef{LTE}{Long Term Evolution}

\colorlet{darkblue}{blue!50!black}
\colorlet{orange}{red!50!yellow}

\title{Outage Analysis of Downlink URLLC in Massive MIMO systems with Power Allocation}
\author{\IEEEauthorblockN{Alexandru-Sabin Bana$^1$, Luca Sanguinetti$^2$, Elisabeth~De~Carvalho$^1$, and Petar Popovski$^1$}
\IEEEauthorblockA{$^1$Department of Electronic Systems, Aalborg University, Denmark \\
$^2$Dipartimento di Ingegneria dell'Informazione, University of Pisa, 56122 Pisa, Italy \\
Email: $^1$\{asb, edc, petarp\}@es.aau.dk, $^2$luca.sanguinetti@unipi.it} \\
}

\begin{document}
\maketitle
\begin{abstract}
Massive MIMO is seen as a main enabler for low-latency communications, thanks to its high spatial degrees of freedom. The channel hardening and favorable propagation properties of Massive MIMO are particularly important for multiplexing several URLLC devices. 
However, the actual utility of channel hardening and spatial multiplexing is dependent critically on the accuracy of channel knowledge. 
%Channel hardening makes fading behave as deterministic, which is critical for achieving high reliability. Favorable propagation makes the channels of different devices asymptotically orthogonal, allowing for more efficient spatial multiplexing. 
%In turn, the efficiency of spatial multiplexing relies on the accuracy of channel knowledge. 
%For the case of several multiplexed devices and low-latency, the training cost becomes critical, and orthogonal pilot allocation within a cell may not be suitable. 
When several low-latency devices are multiplexed, the cost for acquiring accurate knowledge becomes critical, and it is not evident how many devices can be served with a latency-reliability requirement and how many pilot symbols should be allocated. % within a given time-frequency resource limit.
This paper investigates the trade-off between achieving high spectral efficiency and high reliability in the downlink, by employing various power allocation strategies, for maximum ratio and minimum mean square error precoders.
The results show that using max-min SINR power allocation achieves the best reliability, at the expense of lower sum spectral efficiency.
%The effect of a larger number of pilot symbols is also investigated from the spectral efficiency and reliability perspective.
%This paper considers the effect of intra-cell pilot contamination due to non-orthogonal training sequences on the reliability of low-latency communications between a base station and several simultaneously active devices. 
%The final paper will present the trade-off arising between the number of multiplexed devices and the channel training overhead, for several channel estimation and beamforming techniques relying on both instantaneous and long term statistics of the channel.
\end{abstract}

\section{Introduction}
%\subsection{State of the art}
%Massive MIMO is considered to be the enabling wireless technology of the new 5G standard \cite{five_disruptive_tech_5g,massive_mimo_next_generation}.
A massive MIMO system is comprised of a base-station (BS) equipped with a very large number of antennas $M$, which are controlled in a fully digital manner, in order to spatially multiplex a large number of devices simultaneously \cite{Marzetta2010}.
The large number of antennas at the BS provide a tremendous increase in the number of degrees-of-freedom (DoFs), which is paramount in achieving the design goals of 5G \cite{massiveMIMOnetworks_book,massive_mimo_next_generation,massiveMIMO_maximal_SE,asb_maMIMO_MTC}.

Massive MIMO is largely considered as the main technology enabling the high data-rates and high spectral efficiency (SE) required by the extended mobile broadband (eMBB) service within 5G \cite{Marzetta2010, massiveMIMO_maximal_SE, massiveMIMOnetworks_book,five_disruptive_tech_5g}.
However, together with the roll-out of 5G, new services have emerged, complementary to eMBB.
These services are tailored to serving machine-type communications (MTC) and, whether we are regarding massive MTC (mMTC) or ultra-reliable low-latency communications (URLLC), massive MIMO is considered to be their key enabler \cite{asb_maMIMO_MTC} as well.

The requirements of URLLC are considerably stricter than previous services. According to 3GPP \cite{3gpp_38913}, the typical reliability of a URLLC packet of 32 bytes is defined to be $99.999\%$ within {$\SI{1}{ms}$} latency.
The challenges and potential solutions for achieving the strict URLLC requirements have been largely discussed in the literature \cite{PNSCSTBKKS2018,urllc_tcom,Popovski2018,asb_maMIMO_MTC}, and it is worth mentioning a few notable enablers such as: interface diversity\cite{JJN_interf_diversity_urllc}, network slicing \cite{Popovski2018}, forward error correction (FEC) for code diversity \cite{Wang2016}, coherent and non-coherent detection methods with massive BS arrays \cite{feasibility_large_arrays_urllc,dual_stage_non_coh}, precoding based on instantaneous and long-term statistics \cite{asilomar_urllc_maMIMO}.

Other papers have also investigated the use of various linear precoders, such as maximum ratio transmission (MRT) and zero-forcing, in massive MIMO URLLC for multiplexing simultaneous devices \cite{massive_MIMO_tactile_internet}, albeit without accounting for the channel estimation inaccuracy.
The authors of \cite{Wang2016} have proposed a forward error correction code diversity solution in order to separate pilot-interfering users, by assigning unique user signatures.

Approaches based on stochastic network calculus were adopted in \cite{delay_performance_miso_comm} in order to evaluate the latency-reliability trade-off using delay violation probability.
The analysis shows that increasing the number of antennas is beneficial in reducing the delay violation probability.
The authors of \cite{urllc_mmwave_massivemimo} follow a similar approach, where they model the latency-reliability trade-off with a probabilistic constraint on the queue length at the BS.
Joint optimization of power, bandwidth, and the number of active antennas for a given number of active devices has been studied \cite{energy_eff_RA_maMIMO_URLLC} in order to maximize the energy efficiency of a massive MIMO network, subject to Quality-of-Service (QoS) constraints. 

Power control in a single-cell massive MIMO system has been considered in \cite{power_ctrl_single_cell}, and the spectral efficiency (SE) has been used as the main metric to develop algorithms based on the weighted minimum SE among the users and the weighted sum SE.
Another approach has been taken in \cite{EE_pow_alloc} where the main metric for optimization is the energy efficiency in the downlink (DL), with QoS constraints.
A more recent work \cite{Sanguinetti_Deep_learning_power} extended the max-min SE and max-product SE strategies in \cite{massiveMIMOnetworks_book} by proposing the use of a deep learning approach to predict the optimal power allocation policies from the UE positions.

This paper treats a DL URLLC system, where multiple devices are spatially multiplexed simultaneously.
We expand previous knowledge on the trade-offs in terms of SE with insights on what is the corresponding reliability, and the interplay between the two metrics.
The work investigates power allocation strategies and their impact on the SE and reliability, for the case of imperfect channel estimation. 
The number of orthogonal pilots is also varied, in order to quantify the benefit of having a more accurate estimation versus the loss of DoFs for data transmission.

\section{System model}
\label{sec:system}
\begin{figure}[t!]
    \centering
    \includegraphics[width=0.6\linewidth]{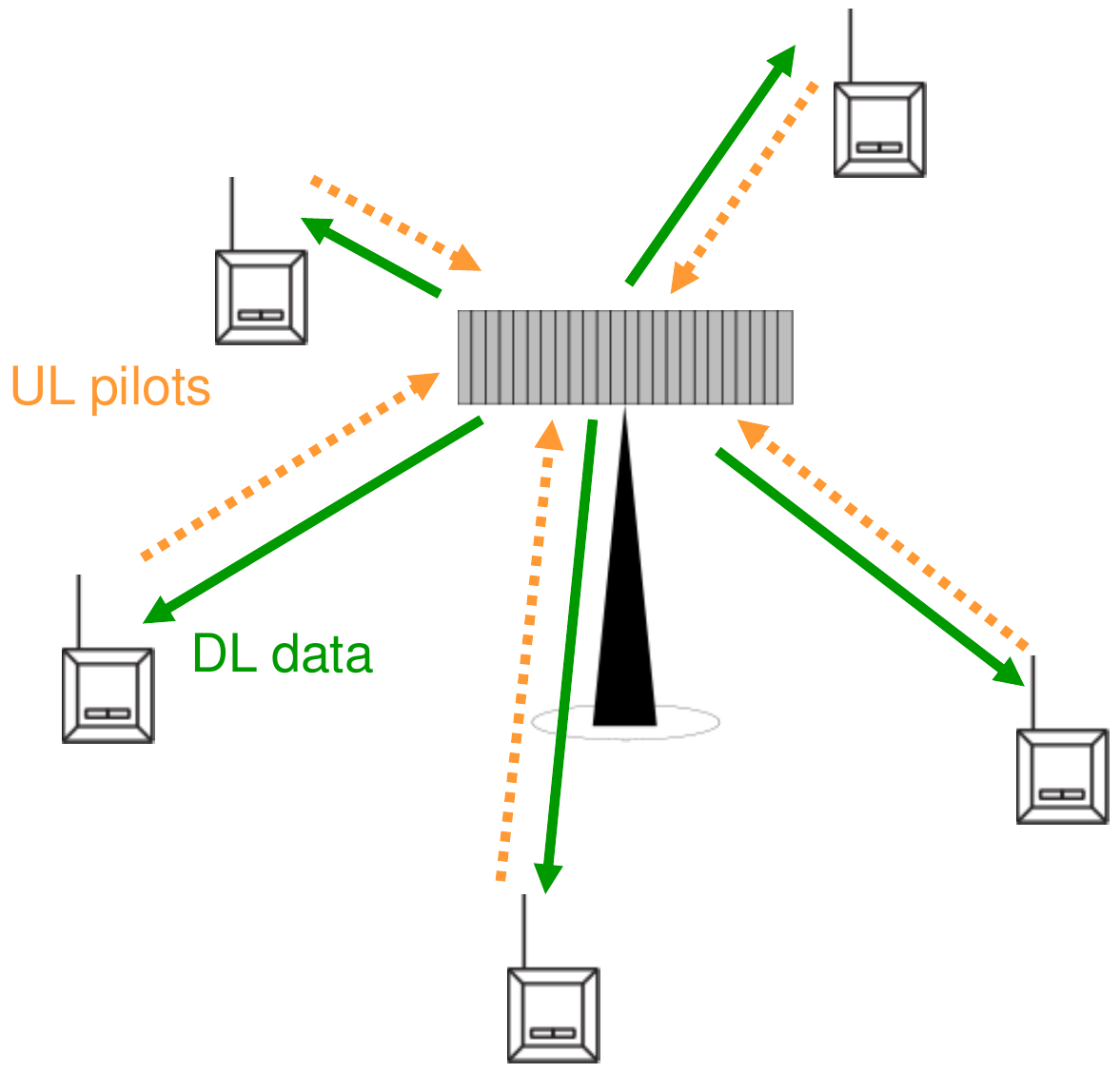}
    \caption{System model of a massive MIMO BS, multiplexing several devices simultaneously in the DL.}
    \label{fig:sytem_model}
    \vspace{-10pt}
\end{figure}

We consider the DL of a single-cell massive MIMO system where the BS is equipped with $M$ antennas and serves $K$ single-antenna devices; see Fig.~\ref{fig:sytem_model}. The $K$ devices need to satisfy the same latency constraint, which is assumed to be lower than the channel coherence time. 

\subsection{Transmission protocol and channel estimation}
We assume that a time-division-duplex (TDD)
protocol is used with a pilot phase for channel estimation, followed by a data transmission phase. We consider the standard block fading TDD protocol \cite[Ch.~2]{massiveMIMOnetworks_book} in which each coherence block consists of $\tau$ channel uses, whereof $\tau_p$ are used for uplink pilots and $\tau_d = \tau- \tau_p$ for downlink data. We assume that $\tau_p = fK$ where the integer $f$ is {the number of pilots per device}. %called pilot reuse factor.
We denote $\mathbf{h}_k$ the channel vector from the BS to UE $k$ and assume that it is modelled as uncorrelated Rayleigh fading, i.e. $\mathbf{h}_k \sim \mathcal{CN}(0,\beta_k\mathbf{I}_M)$, where $\beta_k$ is the large-scale fading coefficient accounting for pathloss and shadowing, defined as \cite[Ch. 2]{massiveMIMOnetworks_book}
\begin{equation}
    \beta_k= \Upsilon -10\alpha \log_{10}\left(\frac{d_k}{\SI{1}{km}}\right) + F_k
    \label{eq:pathloss}
\end{equation}
where $\Upsilon$ is the median channel gain at a reference distance of 1 km from the BS. Both $\Upsilon$ and $\alpha$ are parameters computed from established models \cite{massiveMIMOnetworks_book,3gpp_sim}. $F_k \sim \mathcal{N}(0,\sigma_\text{sf}^2)$ is a random term modeling the shadow fading as a log-normal random variable.

% Low-latency systems operate on the assumption that the latency constraint is lower than the channel coherence time.
%This assumption sets the achievable SE to a considerably smaller level than in broadband systems, due to the limited use of the obtained channel state information.

The uplink pilot sequence of device~$k$ is denoted by ${\bphiu_{k} \in \mathbb{C}^{\taupu}}$ and satisfies $\| \bphiu_{k} \|^2  = \taupu$. The elements of $\bphiu_{k}$ are scaled by the pilot power $\sqrt{p}$ and transmitted over $\taupu$ channel uses. The MMSE estimate of $\vect{h}_{k}$ is \cite[Ch.~3]{massiveMIMOnetworks_book}
\begin{align}
\hat{\mathbf{h}}_k &= \frac{\beta_k}{\beta_k + \frac{1}{\tau_p}\frac{\sigma^2}{ p}} \left(\vect{h}_{k} + \frac{1}{\tau_p\sqrt{p}}\vect{N}^{p}\bphiu_{i}^*\right)\end{align}
where $\vect{N}^{p} \in \mathbb{C}^{M \times \taupu}$ is noise with i.i.d.\ elements distributed as $\CN(0,\sigma^{2})$ and $\sigma^2=-174 + 10\log_{10}(B)+\text{NF} \text{ [dBm]}$, with $\text{NF}$ being the noise factor of the BS. The estimation error $\tilde{\mathbf{h}}_k=\mathbf{h}_k-\hat{\mathbf{h}}_k$ has correlation matrix $\mathbf{C}_k=\beta_k \big(1 - {\beta_k}(\beta_k + \frac{1}{\tau_p}\frac{\sigma^2}{ p})^{-1}\big){\mathbf{I}_M}$ \cite{massiveMIMOnetworks_book}.

\subsection{Spectral efficiency and precoding design}
The BS transmits the DL signal $
\vect{x} = \sum_{i=1}^{K} \vect{w}_{i} \varsigma_{i}$
where  $\varsigma_{i} \sim \CN(0,\rho_i)$ is the DL data signal intended for device~$i$, assigned to a precoding vector $ \vect{w}_{i} \in \mathbb{C}^{M}$ that determines the spatial directivity of the transmission and satisfies $ \|  \vect{w}_{i} \|^2  =1$ so that $\rho_{i}$ represents the transmit power. An achievable DL SE for device $k$ can be computed in Massive MIMO by using the \emph{hardening bound} \cite{Marzetta2010}. This yields 
\begin{equation} \label{eq:downlink-SE-expression-forgetbound}
{\mathsf{SE}}_{k} = \frac{\tau - \tau_p}{\tau} \log_2   (  1 +
\gamma_{k} ) \quad \textnormal{[bit/s/Hz]} 
\end{equation}
with 
\begin{eqnarray}
{\gamma}_k = \dfrac{\rho_k \abs{\mathbb{E}\left[{\mathbf{w}_k^H\mathbf{h}_k}\right]}^2}{\sum_{i=1}^K \rho_i\mathbb{E}\left[\abs{\mathbf{w}_i^H \mathbf{h}_k}^2 \right] - \rho_k \abs{\mathbb{E}\left[\mathbf{w}_k^H\mathbf{h}_k\right]}^2 + \sigma^2}
\label{eq:SINR}
\end{eqnarray}
where the pre-log factor accounts for the fraction of samples per
coherence block used for DL data. The expectations are computed with respect to the channel realizations. This is not a typical assumption for low-latency transmissions, as the latency constraint can only accommodate one coherence block in time.
However, URLLC is expected to use multiple DoFs in frequency, which enables having multiple channel realizations within the available resources.
It should also be noted that as the array becomes larger, the channel hardening effect occurs for fewer number of channel realizations taken in the expectation. %\PP{Good to mention that the SE is based on selecting the rate according to the average behavior of the channel, but an outage can occur if the instance of the channel is worse than that. We cannot do this for this paper, but, in general, the rate (SE) should be selected according to some tail constraints, not according to the average. } %asb: i have included a sentence to clarify the averaging later: This maximum achievable rate is computed using the hardened DL SINR $\gamma_k$ over the multiple DoFs in frequency in \eqref{eq:SINR}.

The achievable SE in \eqref{eq:downlink-SE-expression-forgetbound} holds true for any precoding scheme.
In this work, we select $\vect{w}_{k}$ as $\vect{w}_{k} = { \vect{v}_{k} }/{ \|  \vect{v}_{k} \| } $ with $\vect{v}_{k}$ being designed according to MR and MMSE precoding \cite[Ch.~4]{massiveMIMOnetworks_book}
%\begin{equation} \label{eq:precoding-schemes}
%\vect{v}_{k}^{\rm MR}=\hat{\vect{h}}_{k}
%\end{equation}
%and MMSE combining
%\begin{equation} \label{eq:precoding-schemes2}
%\vect{v}_{k}^{\rm MMSE} = \Bigg(  \sum\limits_{i=1}^K \hat{\vect{h}}_{i} {\hat{\vect{h}}_{i}}^{\Htran} + \sum\limits_{i=1}^{K} \vect{C}_{i}+  \frac{\sigma^2}{p}  \vect{I}_{M} \Bigg)^{\!-1}  \!\!\!  \hat{\vect{h}}_{k}.
%\end{equation}
\begin{equation}
\vect{v}_{k} = \begin{cases}
\hat{\vect{h}}_{k} & \textrm{with MR}, \\
\Bigg(  \sum\limits_{i=1}^K \hat{\vect{h}}_{i} {\hat{\vect{h}}_{i}}^{\Htran} + \sum\limits_{i=1}^{K} \vect{C}_{i}+  \frac{\sigma^2}{p}  \vect{I}_{M} \Bigg)^{\!-1}  \!\!\!  \hat{\vect{h}}_{k}. & \textrm{with MMSE}.
\end{cases}
\end{equation}
This choice is motivated by the fact that MMSE is optimal but has high computational complexity. On the other hand, MR is suboptimal but has the lowest complexity among the receive combining schemes.

%In this paper, we abstract from considering the diversity benefit of the DoFs in the frequency bandwidth $B$, in order to better exemplify the effect of power allocation strategies and the trade-offs occurring when spatially multiplexing URLLC devices.
%We do note that considering several DoFs in frequency in the model is a realistic and interesting assumption, which will be considered as possible extensions of the current paper, as it may introduce further trade-offs and the potential use of scheduling algorithms.
% The effect of frequency diversity is well investigated \ASB{ref}, and this paper takes this simplifying assumption in order to better exemplify the effect of power allocation strategies and the trade-offs occurring when multiplexing URLLC devices.
% The effects of trading off the DoFs in frequency are considered for future extensions of the work.

\section{Problem Formulation}

Several trade-offs arise in the above system model when URLLC requirements are present.
The first is about how many channel uses $\tau_p$ should be utilized as pilots for channel estimation for a given coherence block of $\tau$ channel uses. The more are used for pilots, the fewer are left for data transmission. This inevitably increases the communication latency or decreases the reliability of the data transmission phase.
%The optimal $\tau_p$ depends on the total number of channel uses within the latency-coherence constraint, and the number of active devices $K$. 
Another trade-off comes from the fact that employing power control strategies for increased fairness will lead to lower sum SE in the network.
Ideally, in URLLC one would desire to first fulfill the outage requirements, then to achieve the best possible sum SE of the network.
%However, fairness will improve the minimal SE among users.

The goal of this paper is to investigate how to control the trade-off between SE and reliability by employing power control and dimensioning the pilot size, for the MR and MMSE precoders.

\subsection{Outage Definition}
We assume that in each latency-coherence block the data transmission rate per device is fixed to ${\frac{b}{\tau-\tau_p}}$ measured in bit per channel use. By multiplying it with the coherence bandwidth $B_c$, we obtain the threshold rate $R_T= B_c\frac{b}{\tau-\tau_p} \left[\text{bit}/\text{s}\right]$.
%\asb{in order to be able to deliver the $b$ bits within the latency requirement $\tau$, given that $\tau_p$ symbols are spent on UL pilots}.
Using \eqref{eq:downlink-SE-expression-forgetbound}, the maximum achievable rate for device $k$ over the communication bandwidth $B$ is
\begin{eqnarray}
{R}_{k}= B \frac{\tau - \tau_p}{\tau} \log_2   (  1 +
\gamma_{k} ) \quad \textnormal{[bit/s]}.
\label{eq:outage_rate}
\end{eqnarray}
This maximum achievable rate is computed using the hardened DL SINR $\gamma_k$ over the multiple DoFs in frequency in \eqref{eq:SINR}.
The outage probability of device $k$ measures the probability that the DL SINR $\gamma_k$ cannot support the transmission of the fixed packet of $b$ bits within the latency constraint. It is defined as
\begin{align}
{\Pr}_\text{out}^\text{device} &= \Pr[{R}_{k}<R_T]  \nonumber \\ 
&=\Pr\left[B \frac{\tau - \tau_p}{\tau} \log_2(1+\gamma_k) < B_c\dfrac{b}{\tau-\tau_p}\right]
\label{eq:P_out_dev}
\end{align}
and it is evaluated across the device positions.

In addition to the device outage, we can define the system outage, which is the probability that at least one device of a given setup would be in outage. 
This corresponds to measuring the worst user outage across the device deployments. Its expression is given by
\begin{align}
{\Pr}_\text{out}^\text{sys} = \Pr\left[B \frac{\tau - \tau_p}{\tau} \log_2(1+\min_{K} \gamma_k) < B_c \dfrac{b}{\tau-\tau_p}\right].
\label{eq:P_out_sys}
\end{align}

\subsection{Power Allocation}
The DL transmit powers $\{\rho_k:\forall k\}$ in \eqref{eq:downlink-SE-expression-forgetbound} need to be selected. To achieve high reliability, it is important to increase the fairness between the devices in the network, meaning that devices located further away from the BS, which experience worse channel conditions, should be allocated more power.
This induces a trade-off, between how much power should be allocated to devices experiencing a weak channel, at the expense of a performance degradation of the network sum SE and of devices with stronger channels. To this extent, several power allocation strategies have been proposed in the literature \cite{massiveMIMOnetworks_book,Sanguinetti_Deep_learning_power,power_ctrl_single_cell,EE_pow_alloc}. The most simple solution is equal power allocation, i.e., $\rho_i = P_{\max}/K$ with $P_{\max}$ being the maximum DL transmit power. Alternatively, two prominent examples are the max-min fairness and max-product SINR strategies, which can be mathematically formalized as follows: 
\begin{align} \label{eq:max-min}
\max_{\{\rho_{k}:\forall k\}} & \quad \, \min_{k} \mathsf{SE}_{k} \\
\textrm{subject to}\,\,\,\,\, & \quad  \sum_{k=1}^{K} \rho_{k} \leq P_{\max} \notag
\end{align}
and the max-product SINR, given by
\begin{align} \label{eq:max-prod}
\max_{\{\rho_{k}:\forall k\}} & \quad \,  \prod_{k=1}^{K} \gamma_{k}  \\
\textrm{subject to}\,\,\,\,\, & \quad  \sum_{k=1}^{K} \rho_{k} \leq P_{\max}. \notag
\end{align}
The max-min fairness in \eqref{eq:max-min} provides complete fairness by only counting the SE achieved by the weakest device in the network. This results in the same SE for all; that is, a device has no benefit of having a good channel condition. This inevitably reduces the sum SE. The max-prod power allocation policy \eqref{eq:max-prod} balances between sum SE and fairness. The impact of these schemes has been mostly studied from the perspective of average network SE.
We extend this result to the previously defined metrics of device outage and system outage, and show the interplay between achieving reliability and SE.

%This paper compares the baseline equal power allocation scheme with two potential strategies for increasing the reliability: max-min SINR and max-product SINR.
%
%
%The utility functions of the investigated schemes can be expressed as:
%\begin{align}
%    U_\text{max-min SINR} &= \min_{k=1:K} \gamma_k ; \\
%    U_\text{max-product SINR} &= \prod_{k=1:K} \gamma_k.
%\end{align}
%

\section{Numerical Analysis}
The spectral efficiency, device outage and system outage are evaluated using Monte-Carlo simulations.
Device deployments are generated based on the scenario defined in Section~\ref{sec:system}.
The simulation parameters are described in Table~\ref{table:param}.
The number of symbols in a coherence block is based on the size of the coherence bandwidth, which here is assumed to be $\SI{100}{kHz}$.
The number of channel instances used in taking the expectation in \eqref{eq:SINR} is given by the number of coherence blocks in the total bandwidth, and is equal to $B/B_c$.

\begin{table}[t!]
\centering
\begin{tabu} to \columnwidth {|c|c|}
\hline
BS antennas & $M=100$  \\ \hline
Setup deployments & $N=10^5$  \\ \hline
Cell size & macro-cell\cite{3gpp_sim}, square, length 500 m  \\ \hline
Numer of pilots per device & $f={1,2}$  \\ \hline
Packet size & $b=256$ bits  \\ \hline
Latency requirement & 1 ms  \\ \hline
Coherence block & $\tau = 100$   \\ \hline
Coherence  bandwidth & $B_c= \SI{100}{kHz}$  \\ \hline
Transmission bandwidth & $B= \SI{20}{MHz}$  \\ \hline
BS noise factor &  $\text{NF}=\SI{7}{dB}$ \\ \hline
BS power constraint & $P_{\max}=\SI{46}{dBm}$ \cite{3gpp_sim} \\ \hline
Device power & $p=\SI{23}{dBm}$ \cite{3gpp_sim} \\ \hline
Median channel gain & $\Upsilon=\SI{-148.1}{dB}$ at 1 km \cite{massiveMIMOnetworks_book} \\ \hline
Pathloss exponent & $\alpha=3.76$ \cite{3gpp_sim} \\ \hline
Shadow fading standard deviation & $\sigma_\text{sf}=7$ (NLOS) \\ \hline
\end{tabu}
\vspace{3pt}
\caption{Simulation parameters.}
\label{table:param}
\vspace{-20pt}
\end{table}

\begin{figure}[t!]
    \centering
    \includegraphics[width=1\linewidth]{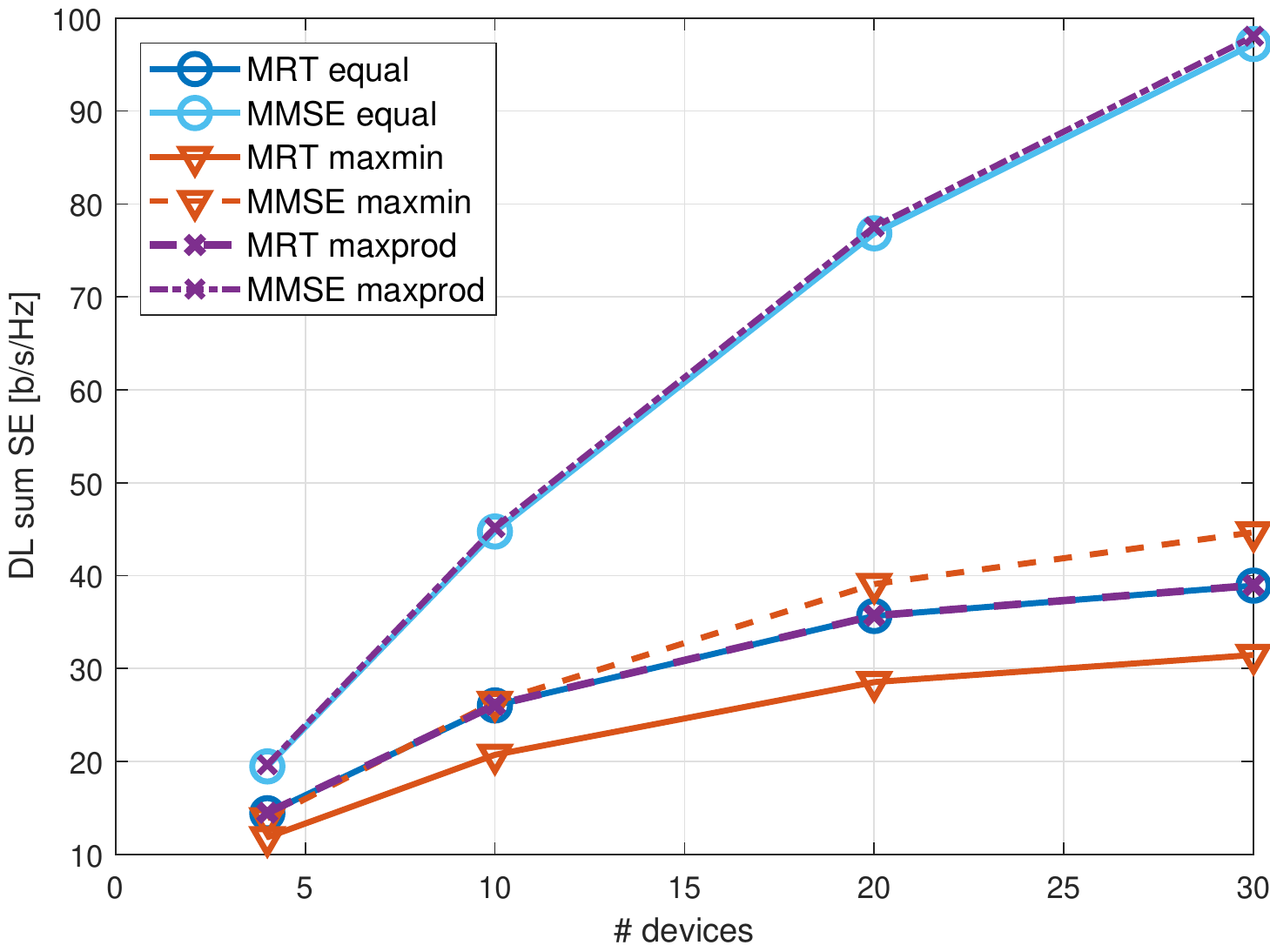}
    \caption{Sum SE comparison, for the different power allocation strategies and MRT and MMSE precoders, for increasing number of devices.}
    \label{fig:SE}
\end{figure}

The network sum SE of the three power allocation strategies for the MRT and MMSE precoders are shown in Fig.~\ref{fig:SE}.
As expected, the sum SE for the max-min SINR is the lowest for each the MRT and MMSE precoders, as it introduces a large amount of fairness in the system, penalizing the SE of strong devices for the benefit of the weak ones.

It is interesting to observe that the max-product SINR method performs nearly identical to the equal power allocation.
This result might seem somewhat surprising, because one would expect the max-product SINR to provide some degrees of fairness.
However, when the CSI is imperfect, the weak devices exhibit worse CSI accuracy, and by allocating more power to inaccurate precoders, the system would create more interference, thereby reducing SINR of other devices and the product of SINRs.
Therefore, max-product SINR, paired with imperfect CSI, performs similarly to equal power. %has a tendency to closely follow the equal power allocation.
%However, let us assume there are two devices only, one with a strong channel and one with a weak channel. 
%If they would have ideal CSI and the interference to each other can be completely cancelled, the max-product SINR would ensure that the devices experience the same SINR.
%However, the weaker one will have a more degraded CSI estimate, and would, therefore, be penalized in terms of SE.

\begin{figure}[t]
    \centering
    \includegraphics[width=1\linewidth]{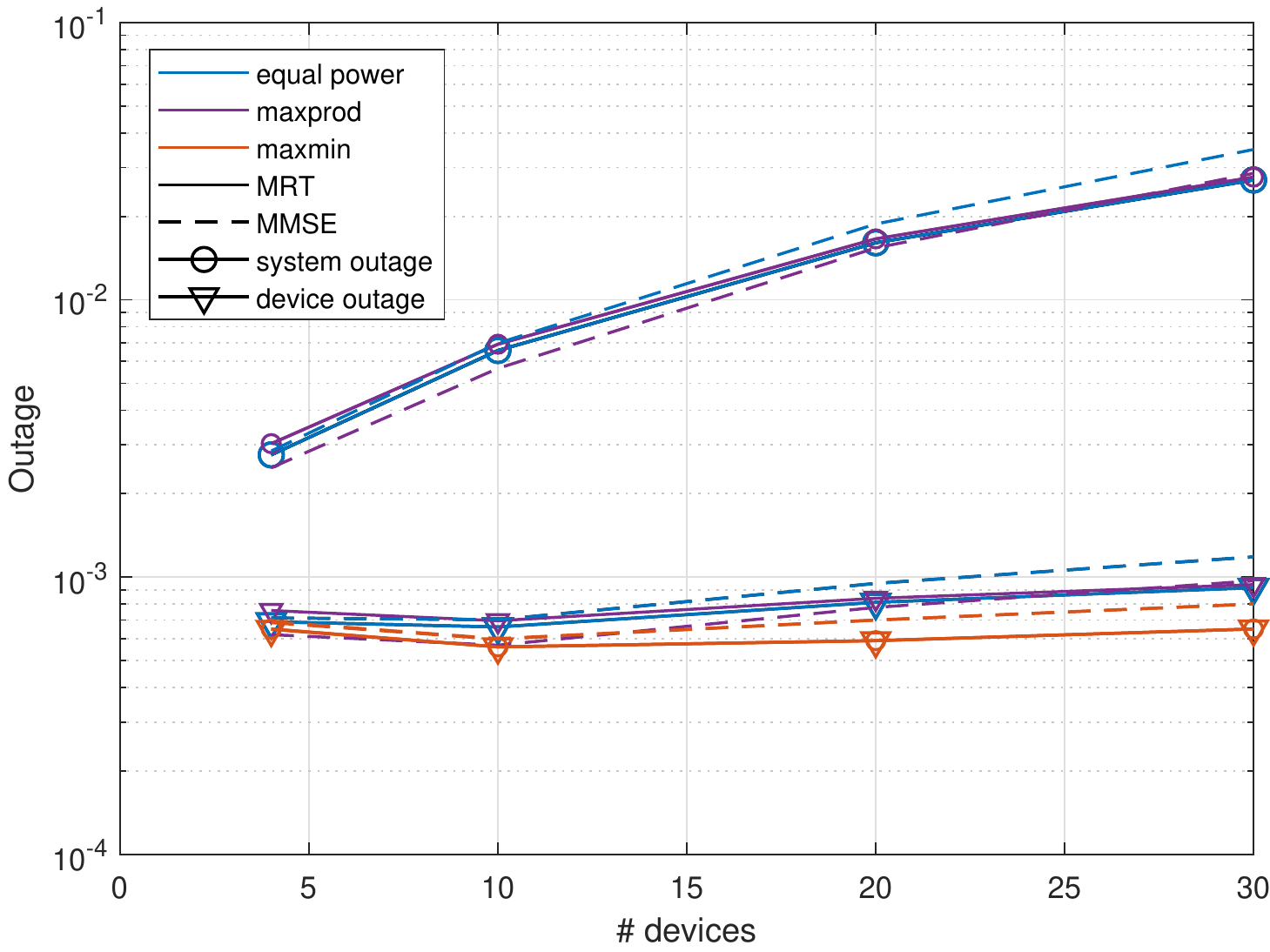}
    \caption{Outage comparison between equal power, max-product SINR, and max-min SINR, for MRT and MMSE.}
    \label{fig:outage1ppd}
\end{figure}
The outage is shown in Fig.~\ref{fig:outage1ppd}.
Firstly, it can be again noticed that the max-product SINR strategy performs nearly the same as equal power allocation, achieving almost identically device and system outage, at an increased complexity of computation.
Secondly, the improvements in fairness of the max-min SINR strategy result in a tremendous decrease of the system outage, which is crucial for URLLC.
Moreover, the device outage is also decreased by a small margin, compared to equal power and max-product SINR.
%This result shows that max-min SINR not only improves the system reliability by a tremendous amount, which is crucial for URLLC, but it also improves the device reliability.
%It can be seen that both the system and the device outage are considerably improved.

Based on the SE and outage results from Fig.~\ref{fig:SE}-\ref{fig:outage1ppd}, it can be noticed that the precoder choice of MMSE over MRT has a significant improvement on the sum SE, while bringing only a minor detriment to the outage.
Moreover, the results show that the sum SE needs to be sacrificed in order to obtain the considerable improvement in outage provided by the max-min SINR strategy, if the system is to operate at its peak reliability.

\begin{figure}[t]
    \centering
    \includegraphics[width=1\linewidth]{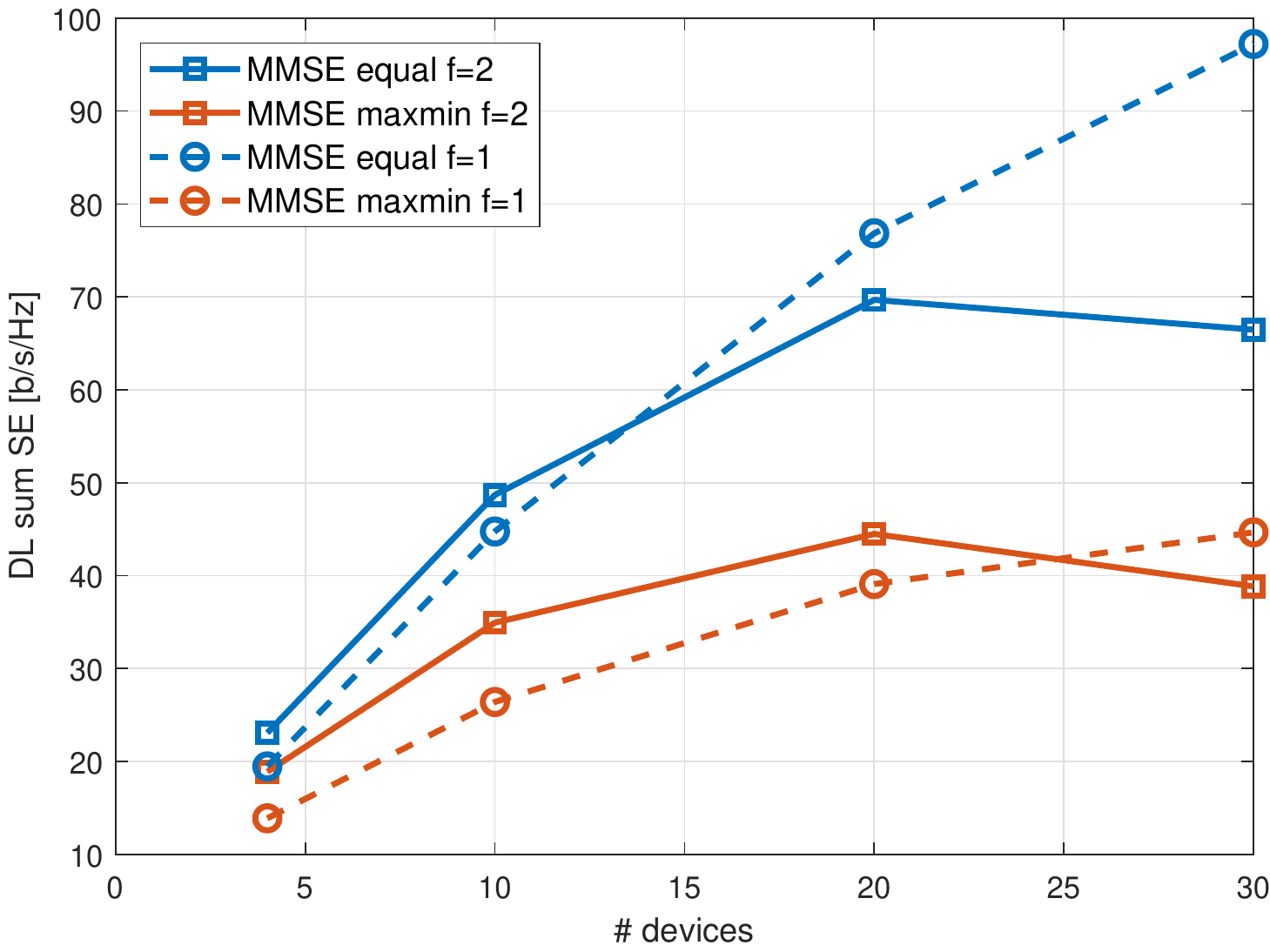}
    \caption{Comparison of the sum SE, for MRT with equal power and max-min SINR power allocation strategies, and with $f=\{1,2\}$ pilots per device.}
    \label{fig:2ppd_se}
\end{figure}

\begin{figure}[t]
    \centering
    \includegraphics[width=1\linewidth]{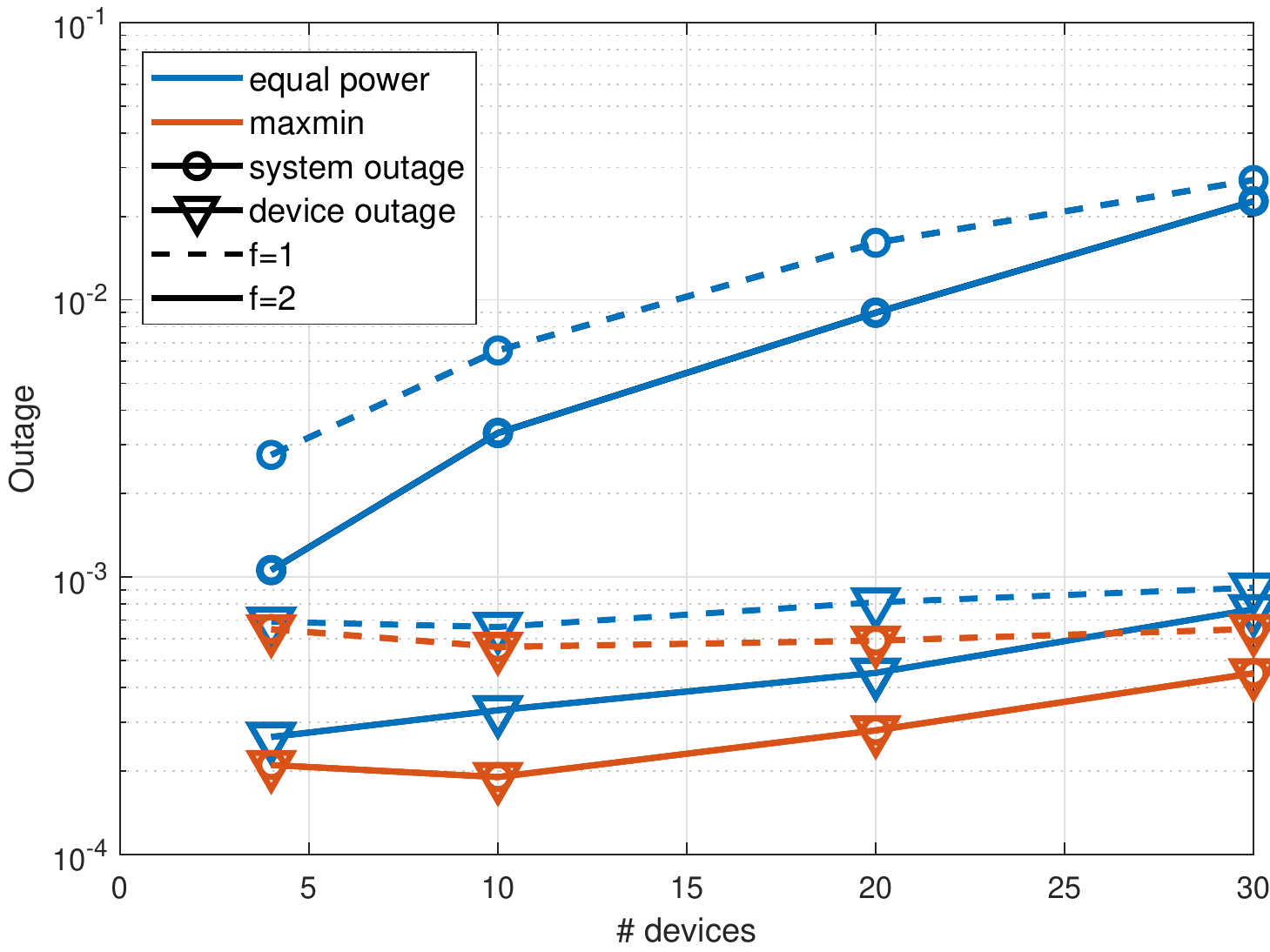}
    \caption{Outage comparison, for MRT with equal power and max-min SINR power allocation strategies, and with $f=\{1,2\}$ pilots per device.}
    \label{fig:2ppd_out}
\end{figure}

Fig.~\ref{fig:2ppd_se} shows the decrease in sum SE when employing two pilots per device for CSI, as the number of devices grows. Fig.~\ref{fig:2ppd_out} shows that from the outage perspective, investing an extra pilot per device for CSI provides further improvements.

\begin{figure}[t]
    \centering
    \includegraphics[width=1\linewidth]{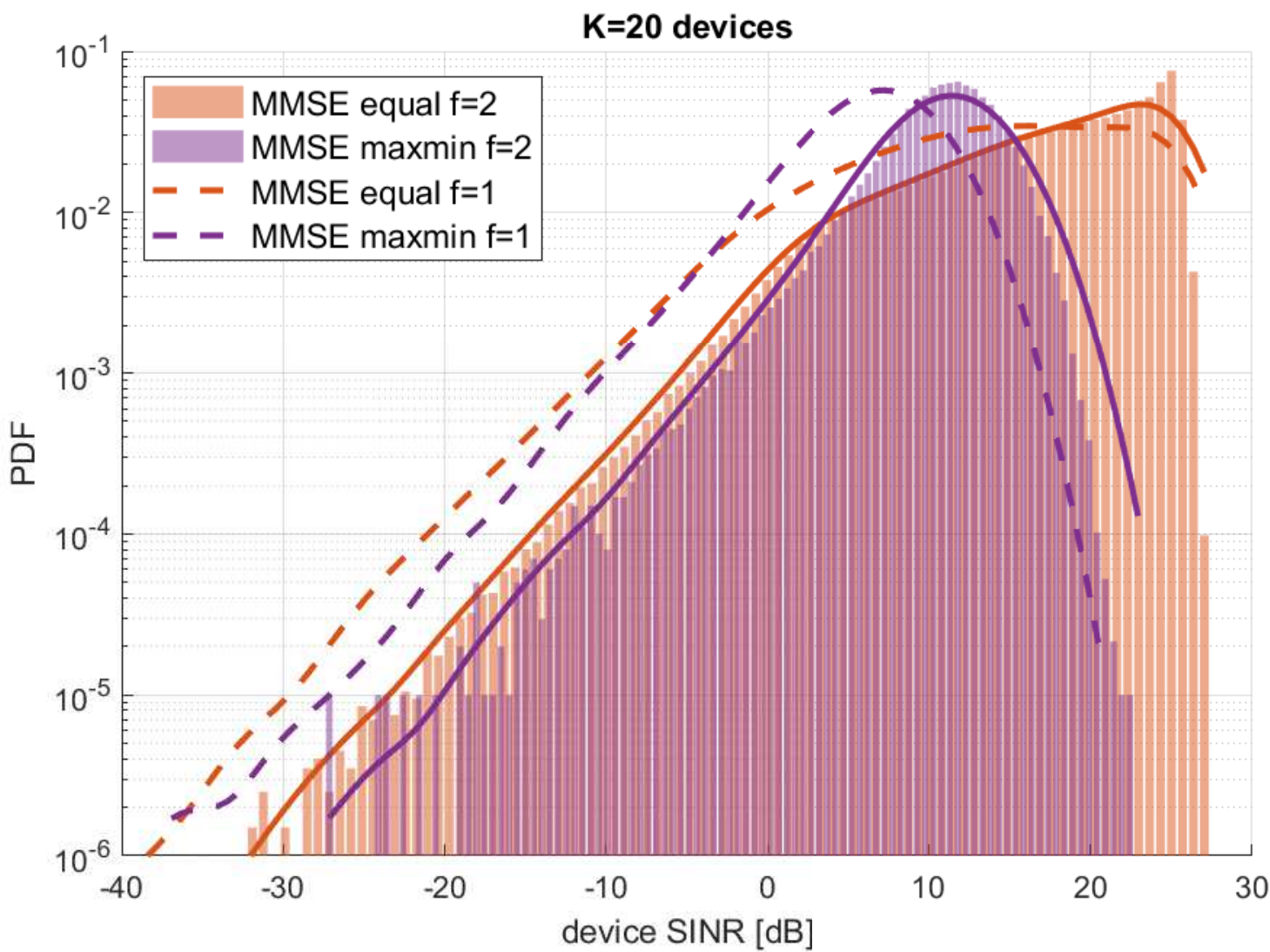}
    \caption{PDFs of the SINRs for MMSE, with equal power and max-min SINR allocation.}
    \label{fig:pdf_mmse}
    \vspace{-10pt}
\end{figure}

In addition, we show the shape of the probability density functions of the device SINRs (${\gamma}_k$) in Fig.~\ref{fig:pdf_mmse}.
One first observation is that the distribution of the max-min SINRs has a smaller lower tail compared to the equal power, which is why the outage is also lower.
%This shows the effect of the power allocation strategy, observed over a large number of setup deployments.
Secondly, it can be seen that the max-min SINR strategy does not achieve the same high SINRs as equal power, due to the fairness. This is the reason why the sum SE of max-min experiences a dramatic decrease (Fig.~\ref{fig:2ppd_se}, note that the X-axis of the SINR is in dB).
%Moreover, it can be observed by comparing the PDF of the max-min SINR strategy for MRT and MMSE, that the MMSE has slightly larger lower tail than MRT (meaning slightly higher outage), but considerably larger upper tail (reflected in the considerably higher sum SE).
% Moreover, it can be observed that when employing max-min, the SINR PDF of the MMSE is experiencing a minor degradation in the lower tail (also noticed in the outage in Fig.~\ref{fig:outage1ppd}), but a considerable improvement in the upper SINR values (reflected in the sum SE in Fig.~\ref{fig:SE}).

% \begin{figure}[ht]
%     \centering
%     \includegraphics[width=1\linewidth]{figures/pdf_maxmin.eps}
%     \caption{PDF of the SINRs for MRT and MMSE with max-min power allocation, with two piots per device. \ASB{i will compress this and the previous figure into 1}}
%     \label{fig:pdf_maxmin}
% \end{figure}

\section{Conclusion}
This paper investigated the means of achieving ultra-reliable low-latency communications in a single-cell downlink massive MIMO system.
More specifically, we utilized state-of-the-art power allocation strategies and precoders, in order to show novel results in terms of outage.
Moreover, we exposed two trade-offs: between achieving high SE and high reliability by power allocation strategies; and between allocating a single uplink pilot symbol versus two pilot symbols per device, for the case of multiplexing several devices in a low-latency setting.

We concluded that the max-min SINR strategy is the best in terms of outage, due to its increased fairness among devices. However, it leads to a much lower sum SE due to allocating more power to the weaker devices. 
The max-product SINR strategy achieves nearly identical performance as the equal power, at an increased complexity of computation, making it the less suitable scheme in this scenario with imperfect CSI.

Furthermore, the results showed that the precoder choice is not highly important for the reliability, but choosing MMSE can considerably improve the sum SE with a minimal decrease in reliability.

The use of an extra pilot symbol per device proved to be beneficial in terms of outage for this particular low-latency setting, despite the decrease in terms of sum SE for the case of larger number of devices.

%\bibliographystyle{IEEEtran}
%\bibliography{mybib}

\section*{Acknowledgment}
The researchers from Aalborg University have been supported partly by the European Research Council (ERC Consolidator Grant nr. 648382 WILLOW) and by the Danish Council for Independent Research (DFF) (CELEST project).
L. Sanguinetti was supported by Pisa University under the PRA 2018-2019 Project CONCEPT.
\end{document}